\documentclass[runningheads]{llncs}

\usepackage[T1]{fontenc}

\usepackage{graphicx}
\usepackage{listings}
\usepackage{cite}
\usepackage[colorlinks=true, allcolors=blue]{hyperref}
\usepackage{multirow}
\usepackage{amsmath}
\usepackage{amssymb}
\usepackage{mathtools}

\newcommand\Tstrut{\rule{0pt}{2.5ex}}         
\newcommand\Bstrut{\rule[-0.9ex]{0pt}{0pt}}   

\begin{document}
\title{Dual-Branch Squeeze-Fusion-Excitation Module for Cross-Modality Registration of Cardiac SPECT and CT}

\titlerunning{Dual Squeeze-Fusion-Excitation Registration}

\author{Xiongchao Chen\inst{1} \and
Bo Zhou\inst{1} \and
Huidong Xie\inst{1} \and
Xueqi Guo\inst{1} \and
Jiazheng Zhang\inst{2} \and
Albert J. Sinusas\inst{1, 2, 3} \and
John A. Onofrey\inst{1,2} \and
Chi Liu\inst{1,2}}

\authorrunning{X. Chen et al.}

\institute{Department of Biomedical Engineering, Yale University, USA \and
Department of Radiology and Biomedical Imaging, Yale University, USA \and
Department of Internal Medicine, Yale University, USA \\
\email{\{xiongchao.chen, chi.liu\}@yale.edu}}

\maketitle  

\begin{abstract}
Single-photon emission computed tomography (SPECT) is a widely applied imaging approach for diagnosis of coronary artery diseases. Attenuation maps ($\mu$-maps) derived from computed tomography (CT) are utilized for attenuation correction (AC) to improve diagnostic accuracy of cardiac SPECT. However, SPECT and CT are obtained sequentially in clinical practice, which potentially induces misregistration between the two scans. Convolutional neural networks (CNN) are powerful tools for medical image registration. Previous CNN-based methods for cross-modality registration either directly concatenated two input modalities as an early feature fusion or extracted image features using two separate CNN modules for a late fusion. These methods do not fully extract or fuse the cross-modality information. Besides, deep-learning-based rigid registration of cardiac SPECT and CT-derived $\mu$-maps has not been investigated before. In this paper, we propose a Dual-Branch Squeeze-Fusion-Excitation (DuSFE) module for the registration of cardiac SPECT and CT-derived $\mu$-maps. DuSFE fuses the knowledge from multiple modalities to recalibrate both channel-wise and spatial features for each modality. DuSFE can be embedded at multiple convolutional layers to enable feature fusion at different spatial dimensions. Our studies using clinical data demonstrated that a network embedded with DuSFE generated substantial lower registration errors and therefore more accurate AC SPECT images than previous methods. Our source code is available at \href{https://github.com/XiongchaoChen/DuSFE_CrossRegistration}{https://github.com/XiongchaoChen/DuSFE\_CrossRegistration}.

\end{abstract}

\section{Introduction}
Myocardial perfusion imaging (MPI) using single-photon emission computed tomography (SPECT) is the most widely performed nuclear cardiology exam for diagnosis of ischemic heart diseases \cite{danad2017comparison}. Photon attenuation within the patient's body is the major limitation that affects accurate diagnosis of cardiac SPECT \cite{singh2007attenuation}. In clinical practice, attenuation maps ($\mu$-maps) derived from computed tomography (CT) are utilized for attenuation correction (AC) \cite{tavakoli2019quantitative}, which largely increases the diagnostic accuracy \cite{patchett2017does}. However, SPECT and CT scans are usually performed sequentially. Thus, improper patient positioning, voluntary movements, or mechanical misalignment of scanners could potentially induce the misregistration between SPECT and CT \cite{goetze2007prevalence}. Previous studies showed that the misregistration can produce errors in the activity distribution of AC SPECT images \cite{fricke2004method}. Thus, cross-modality registration of SPECT/CT images is of vital importance for accurate clinical diagnosis of cardiac SPECT. 

Cross-modality registration is quite challenging because the same structures in two modalities could present totally different intensity and texture patterns \cite{barbu2009boosting}. Traditional methods for cross-modality registration are based on either minimizing the absolute errors between image intensities within overlapping regions \cite{gerlot1992registration} or maximizing the mutual information (MI) between image volumes \cite{maes1997multimodality}. However, these metric-based methods typically ignore the image spatial information, and thus show poor performance in the registration of SPECT and CT-derived $\mu$-maps, since some structures could be presented in the CT transmission scan but missing in the SPECT emission scan.

Deep learning has shown great potential in medical image registration \cite{chen2021deep}. Many methods using convolutional neural networks (CNN) were developed for cross-modality image registration. Sun \textit{et al.} \cite{sun2018towards} proposed a DVNet to estimate the displacement vectors between CT and Ultrasound (US) images, in which image features were extracted by two separate CNN streams and channel-wise concatenated as a late fusion. Xu \textit{et al.} \cite{xu2020adversarial} proposed a translation-based network for deformable registration of CT and magnetic resonance (MR) images, in which uni-modality registration of MR and CT-translated MR was combined with cross-modality registration of MR and CT. Guo \textit{et al.} \cite{guo2020deep} developed a MSReg for rigid registration of transrectal ultrasound (TRUS) and MR images. The two modalities were first concatenated at the input layer as an early fusion, and then fed to a series of ResNetXt \cite{xie2017aggregated} for deep registration. Song \textit{et al.} \cite{song2021cross} adopted a non-local neural network \cite{wang2018non} to recalibrate voxel weights of image features for rigid registration of TRUS and MR images. 

\begin{figure}[htb!]
\centering
\includegraphics[width=0.95\textwidth]{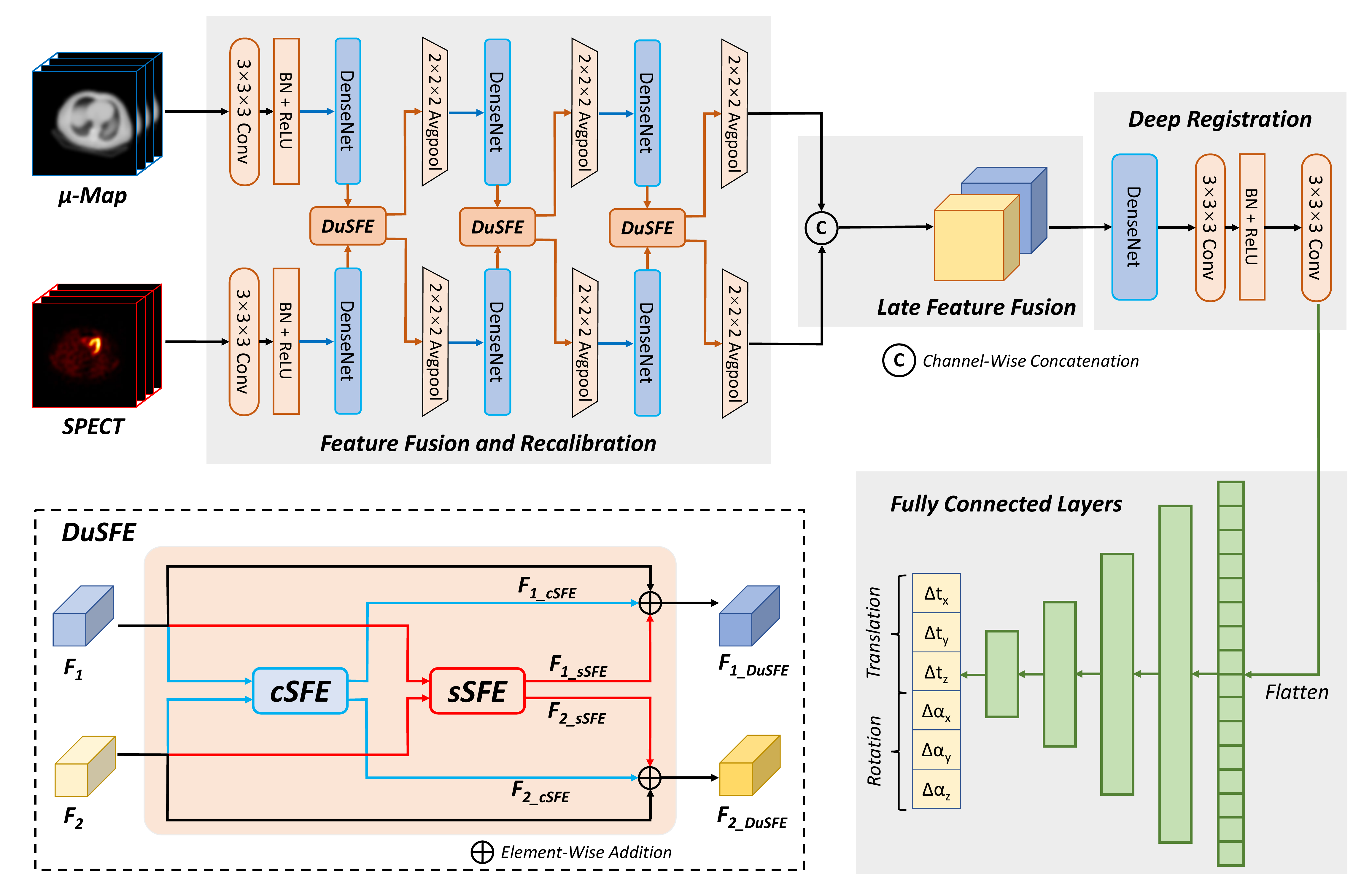}
\caption{Schematic of the proposed registration framework. The input $\mu$-map and SPECT image are fed into two cross-connected CNN streams embedded with DuSFE for feature fusion. The extracted features are concatenated and input to a deep registration module followed by fully connected layers to estimate registration parameters.}
\label{fig:overview}
\end{figure}

However, the aforementioned deep learning methods do not fully extract or fuse the cross-modality information. Some methods directly concatenate the two input modalities as an early fusion \cite{xu2020adversarial, guo2020deep}. Others first extract image features using two separate CNN streams and then concatenate the extracted features for a late fusion \cite{sun2018towards, song2021cross}. The extracted features can also be fused at different downsampling layers as recursive late fusion \cite{liu2020contrastive}. In addition, deep-learning-based rigid registration of cardiac SPECT and CT-derived $\mu$-maps has not been investigated before. In this study, we propose a novel Dual-Branch Squeeze-Fusion-Excitation (DuSFE) attention module for feature fusion, recalibration, and registration of cardiac SPECT and CT-derived $\mu$-maps, inspired by Squeeze-and-Excitation Networks (SENet) \cite{hu2018squeeze}. SENet was developed to recalibrate channel weights of feature maps, which was applied for image fusion \cite{joze2020mmtm}, segmentation \cite{roy2018concurrent}, and transformation\cite{chen2021ct, chen2022direct}. In our study, the DuSFE module fuses the information of multiple modalities and then recalibrates both the channel-wise and spatial features of each modality in a dual-branch manner. DuSFE can be further embedded at different downsampling layers, enabling gradual feature fusion at different spatial dimensions. Our patient studies demonstrated that a network built with DuSFE generated significantly lower registration errors and more accurate AC SPECT images with a minimal increase in computational cost compared to previous methods.

\section{Methods}
\subsection{Dataset and Preprocessing}
The dataset of this study consisted of 450 anonymized clinical cardiac $^{\mathrm{99m}}$Tc-tetrofosmin SPECT MPI studies acquired on GE NM/CT 850c, a hybrid dual-head SPECT/8-slice CT scanner. The SPECT image with a matrix size of $64\times64\times64$ and a voxel size of $6.8\times6.8\times6.8\mathrm{mm}^3$ was reconstructed from 60-angle projections within photopeak window (126.5 to 154.5 keV) using ordered subset expectation maximization (OSEM, 10 subsets, 12 iterations) and then mean-normalized. The original voxel spacing of the CT images is $0.98\times0.98\times5\mathrm{mm}^3$ with a matrix size of $512\times512$ in each slice. The CT values in Hounsfield unit were transformed into attenuation coefficients of $\mu$-maps with the unit of $\mathrm{cm}^{-1}$. The $\mu$-maps were then cropped and interpolated into the same matrix and voxel size as SPECT images. The CT-derived $\mu$-maps were all manually checked and registered with the SPECT images by technologists using vendor software. No data intensity augmentation was implemented in this study.

In clinical practice of cardiac SPECT, the manual corrections are done only for rigid motions. Thus, we consider rigid motions in this study. A group of randomly generated transformation parameters ($\Delta t_x$, $\Delta t_y$, $\Delta t_z$, $\Delta \alpha_x$, $\Delta \alpha_y$, $\Delta \alpha_z$) were used to transform $\mu$-maps to simulate misregistration. Ranges of translations ($\Delta t_x$, $\Delta t_y$, $\Delta t_z$) and rotations ($\Delta \alpha_x$, $\Delta \alpha_y$, $\Delta \alpha_z$) were limited to (8, 8, 4) voxels and (10, 10, 30) degrees with the interval of 0.01 voxels and 0.01 degrees. Wider ranges were assigned for $\Delta t_x$, $\Delta t_y$, and $\Delta \alpha_z$ since misregistration of SPECT and CT mainly exists within the transverse plane. Two groups of parameters were generated for each patient study. In total, 400, 100, and 400 cases were generated for training, validation, and testing.

\subsection{Dual-Branch Squeeze-Fusion-Excitation Module}

\begin{figure}[htb!]
\centering
\includegraphics[width=0.95\textwidth]{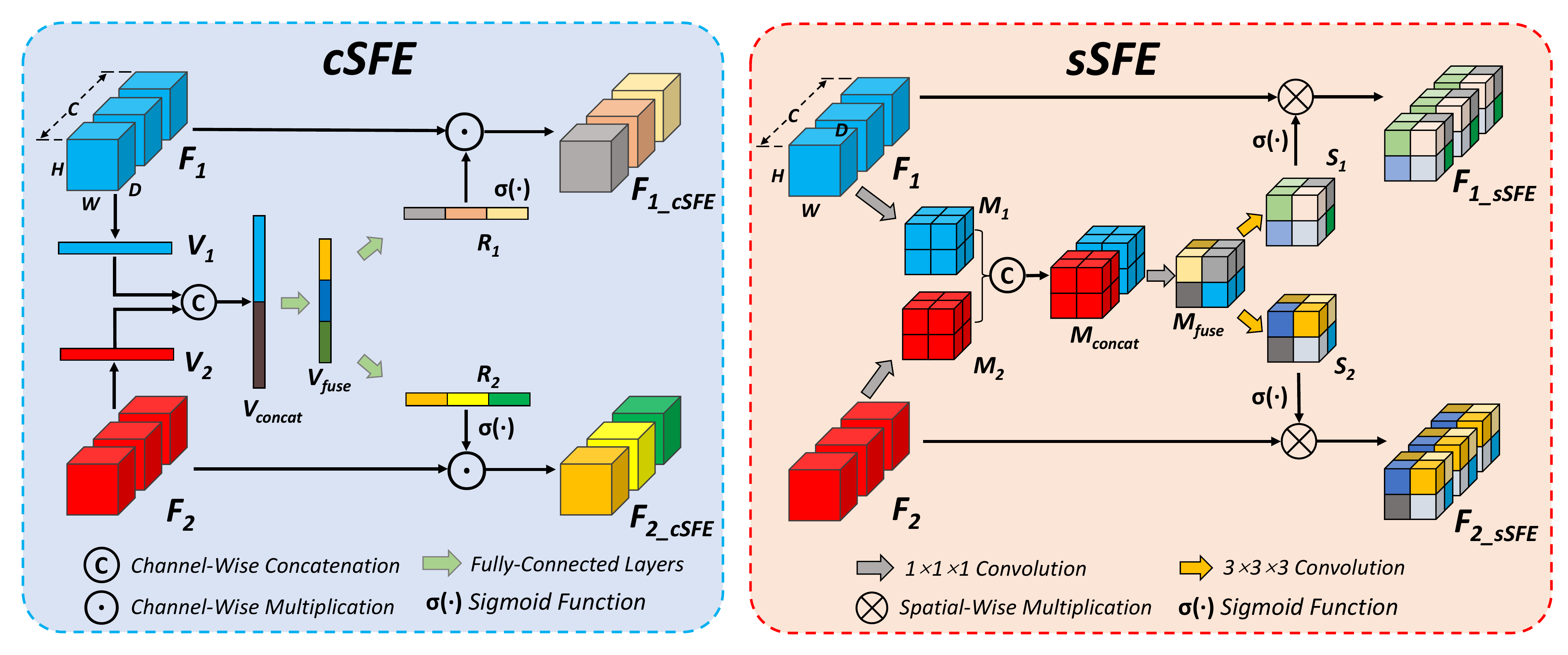}
\caption{Frameworks of cSFE and sSFE attention modules. The channel-wise or spatial features are encoded, fused, and applied back for recalibration in cSFE or sSFE.}
\label{fig:dusfe}
\end{figure}

\noindent The overview of our proposed registration network is presented in Fig.~\ref{fig:overview}. The $\mu$-map and SPECT image are input into two cross-connected CNN streams for feature fusion. Each CNN stream has a 3-layer downsampling structure. In each layer, two modalities are first fed into DenseNet \cite{huang2017densely} for feature extraction, and then jointly input to DuSFE for feature fusion and recalibration.

The overview of our proposed DuSFE module is presented in the bottom dash box of Fig.~\ref{fig:overview}. Two modalities are jointly input into two attention modules, channel-Squeeze-Fusion-Excitation (\textbf{cSFE}) for channel-wise recalibration and spatial-Squeeze-Fusion-Excitation (\textbf{sSFE}) for spatial recalibration. The frameworks of cSFE and sSFE attention modules are presented in Fig.~\ref{fig:dusfe}

In \textbf{cSFE}, two modalities $F_1$ and $F_2 \in \mathbb{R}^{H \times W \times D \times C}$ are squeezed into two vectors $V_1$ and $V_2 \in \mathbb{R}^{C}$ using global average pooling:
\begin{equation}
    V_1(c) = \frac{1}{H \times W \times D} \sum^H_i \sum^W_j \sum^D_k F_1(i,j,k,c),
\end{equation}
\begin{equation}
    V_2(c) = \frac{1}{H \times W \times D} \sum^H_i \sum^W_j \sum^D_k F_2(i,j,k,c),
\end{equation}
where $c$ refers to the $c^{th}$ channel of $F_1$ and $F_2$. Then, $V_1$ and $V_2$ are channel-wise concatenated and fed into a fully connected layer with a weight of $w \in \mathbb{R}^{C \times 2C}$ and a bias of $b \in \mathbb{R}^{C}$ to produce a fused vector $V_{fuse} \in \mathbb{R}^{C}$ for feature fusion:
\begin{equation}
    V_{fuse} = w [V_1, V_2] + b,
\end{equation}
where $[\cdot]$ refers to the channel-wise concatenation operator. Next, $V_{fuse}$ is input to two separate fully connected layers to generate two channel-wise recalibration weights $R_1$ and $R_2 \in \mathbb{R}^{C}$:
\begin{equation}
    R_1 = w_1 V_{fuse} + b_1, R_2 = w_2 V_{fuse} + b_2,
\end{equation}
where $w_1$ and $w_2 \in \mathbb{R}^{C \times C}$, $b_1$ and $b_2 \in \mathbb{R}^{C}$ are the weights and biases of the fully-connected layers. $R_1$ and $R_2$ are then applied back to the two input modalities using channel-wise multiplication, generating $F_{1\_cSFE}$ and $F_{2\_cSFE}$:
\begin{equation}
    F_{1\_cSFE} = \sigma(R_1) \odot F_1, F_{2\_cSFE} = \sigma(R_2) \odot F_2,
\end{equation}
where $\sigma(\cdot)$ is sigmoid function and $\odot$ refers to the channel-wise multiplication.

In \textbf{sSFE}, $F_1$ and $F_2 \in \mathbb{R}^{H \times W \times D \times C}$ are squeezed using convolutional layers with kernels $K_{in1}$ and $K_{in2} \in \mathbb{R}^{1 \times 1 \times 1 \times 1 \times C}$, generating two volumes $M_1$ and $M_2 \in \mathbb{R}^{H \times W \times D \times 1}$:
\begin{equation}
    M_1 = K_{in1} * F_1, M_2 = K_{in2} * F_2, 
\end{equation}
where $*$ refers to the convolution operator. Then, $M_1$ and $M_2$ are channel-wise concatenated and fed into a convolutional layer with a kernel $K_{fuse} \in \mathbb{R}^{1 \times 1 \times 1 \times 1 \times 2}$ to produce a fused volume $M_{fuse} \in \mathbb{R}^{H \times W \times D \times 1}$:
\begin{equation}
    M_{fuse} = K_{fuse} * [M_1, M_2]. 
\end{equation}
Next, $M_{fuse}$ is input to two separate convolutional layers with kernels $K_{out1}$ and $K_{out2} \in \mathbb{R}^{3 \times 3 \times 3 \times 1 \times 1}$ to generate spatial recalibration weights $S_1$ and $S_2$:
\begin{equation}
    S_1 = K_{out1} * M_{fuse}, S_2 = K_{out2} * M_{fuse}.
\end{equation}
$S_1$ and $S_2$ are then applied back to the two input modalities using spatial multiplication, generating $F_{1\_sSFE}$ and $F_{2\_sSFE}$:
\begin{equation}
    F_{1\_sSFE} = \sigma(S_1) \otimes F_1, F_{2\_sSFE} = \sigma(S_2) \otimes F_2,
\end{equation}
where $\otimes$ refers to the spatial multiplication operator.

Finally, the channel-wise and spatial recalibrated feature maps are combined with the input features using element-wise addition:
\begin{equation}
    F_{1\_DuSFE} = F_1+F_{1\_cSFE}+F_{1\_sSFE}, 
\end{equation}
\begin{equation}
    F_{2\_DuSFE} = F_2+F_{2\_cSFE}+F_{2\_sSFE}.
\end{equation}
With DuSFE, information from two modalities can be effectively fused to recalibrate the channel-wise and spatial features for each modality.

\subsection{Deep Registration and Fully Connected Layers}
As is shown in Fig.~\ref{fig:overview}, image features of $\mu$-map and SPECT recalibrated by DuSFE are then channel-wise concatenated for a late feature fusion. The fused features are input to a deep registration module to extract the spatial registration information. Fully connected layers are finally applied to the flattened volumes, estimating the rigid transformation parameters.

\subsection{Implementation Details}
In this study, we included a traditional method using mutual information \cite{maes1997multimodality}, and recent deep learning methods DVNet \cite{sun2018towards}, MSReg (one-stage) \cite{guo2020deep} and non-local attention \cite{song2021cross} as benchmarks. The DuSFE modules in Fig.~\ref{fig:overview} were removed for ablation study, labeled as DenseNet. DenseNet, DenseNet with non-local attention added, and DenseNet with the proposed DuSFE were tested for comparison. All networks were trained for 300 epochs using Adam optimizers ($\beta_1=0.5, \beta_2=0.99$) \cite{kingma2014adam} with an initial learning rate of $5\times 10^{-5}$ and a decay rate of 0.99 per epoch. All the networks were supervised by $L1$ Loss between the predicted and ground-truth 6-digit transformation parameters. All the frameworks were built using PyTorch \cite{paszke2019pytorch} and trained/tested on an NVIDIA Quadro RTX 8000 graphic card with a batch size of 4.

\subsection{Quantitative Evaluations}
The averaged registration errors of translation ($\Delta T$) and rotation ($\Delta R$) were calculated as:
\begin{equation}
    \Delta T = \left( |\Delta \hat{t}_x - \Delta t_x|_1 + |\Delta \hat{t}_y - \Delta t_y|_1 + |\Delta \hat{t}_z - \Delta t_z|_1 \right)/3,
\end{equation}
\begin{equation}
    \Delta R = \left( |\Delta \hat{\alpha}_x - \Delta \alpha_x|_1 + |\Delta \hat{\alpha}_y - \Delta \alpha_y|_1 + |\Delta \hat{\alpha}_z - \Delta \alpha_z|_1 \right)/3,
\end{equation}
where ($\Delta \hat{t}_x$, $\Delta \hat{t}_y$, $\Delta \hat{t}_z$, $\Delta \hat{\alpha}_x$, $\Delta \hat{\alpha}_y$, $\Delta \hat{\alpha}_z$) and ($\Delta t_x$, $\Delta t_y$, $\Delta t_z$, $\Delta \alpha_x$, $\Delta \alpha_y$, $\Delta \alpha_z$) are predicted and ground-truth parameters. Voxel-wise errors of the registered $\mu$-maps and reconstructed AC SPECT images were quantified using normalized mean squared error (NMSE) and normalized mean absolute error (NMAE). Paired t-tests were applied for statistical evaluations.

\begin{table} [htb!]
\caption{Registration errors of translation (mm) and rotation (degrees), number of network parameters, training time per batch, and testing time per case. The best results are marked in \textbf{bold}.}
\label{tab:reg_errors} 
\scriptsize
\centering
\begin{tabular}{ l | c | c || c | c | c}
\hline
\textbf{Methods}                   & $\boldsymbol{\Delta T}$ \textbf{(mm)}    & $\boldsymbol{\Delta R}$ \textbf{(degrees)}      & $\boldsymbol{\#}$\textbf{Parameter}    & \textbf{Train}  & \textbf{Test}  \Tstrut\Bstrut\\  
\hline 
Baseline (Motion)                      & $22.78 \pm 7.68$                   & $8.12 \pm 3.17$                    & \textendash           & \textendash              & \textendash        \Tstrut\Bstrut\\
Mutual Information \cite{maes1997multimodality}   & $7.96 \pm 5.10$          & $2.80 \pm 2.08$                   & \textendash           & \textendash              & 3\textendash5s     \Tstrut\Bstrut\\
\hline 
DVNet \cite{sun2018towards}            & $4.35 \pm 2.11$                    & $1.29 \pm 0.79$                    & 5,440,764             & 0.48s                    & <0.05s            \Tstrut\Bstrut\\
MSReg \cite{guo2020deep}               & $3.94 \pm 1.84$                    & $1.16 \pm 0.78$                    & 11,820,998            & 0.88s                    & <0.05s            \Tstrut\Bstrut\\
DenseNet \cite{huang2017densely}       & $3.88 \pm 1.77$                    & $1.01 \pm 0.57$                    & 11,215,590            & 1.08s                    & <0.05s            \Tstrut\Bstrut\\
DenseNet+Non-Local Attention \cite{song2021cross} & $3.81 \pm 1.63$         & $0.70 \pm 0.42$                    & 11,419,846            & 1.19s                    & <0.05s            \Tstrut\Bstrut\\
DenseNet+DuSFE (Proposed)              & $\textbf{3.33} \pm \textbf{1.63}$  & $\textbf{0.61} \pm \textbf{0.45}$  & 11,726,397            & 1.25s                    & <0.05s            \Tstrut\Bstrut\\
\hline 
\end{tabular}
\end{table}

\section{Results}
\begin{figure}[htb!]
\centering
\includegraphics[width=1.00\textwidth]{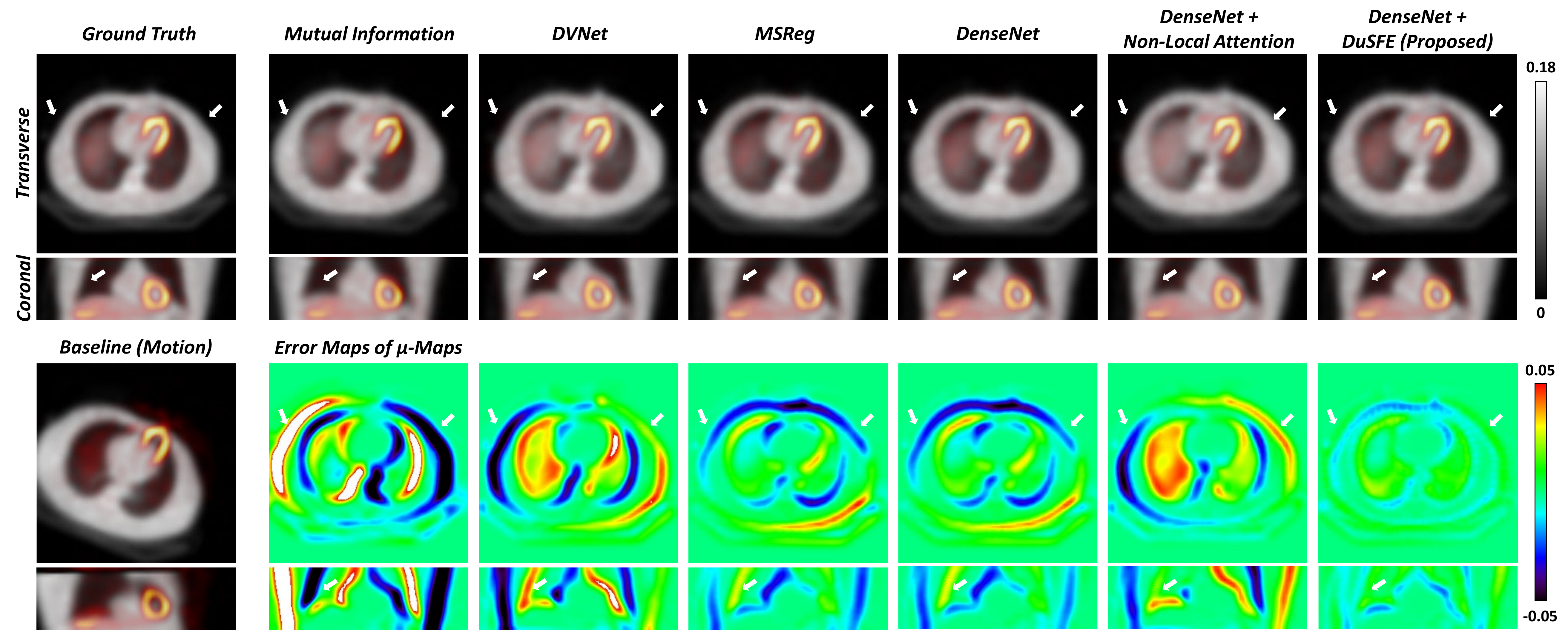}
\caption{Registered $\mu$-maps (unit: $\mathrm{cm}^{-1}$) with overlapped SPECT images using the generated transformation parameters.}
\label{fig:amap}
\end{figure}

As is shown in Table~\ref{tab:reg_errors}, all deep learning methods produced lower registration errors and shorter testing time than Mutual Information. Due to the shallow convolutional layers, DVNet output the lowest accuracy even though consumed fewer computational resources than other deep learning methods. DenseNet outperformed MSReg ($p[\Delta T]$=0.65, $p[\Delta R]$<0.001 with paired t-tests) since densely-connected layers can better extract input information. DenseNet+DuSFE significantly outperformed DenseNet ($p[\Delta T]$<0.001, $p[\Delta R]$<0.001) and DenseNet + Non-Local Attention ($p[\Delta T]$<0.001, $p[\Delta R]$<0.001). This demonstrated that our proposed DuSFE modules improved network performance and showed superior performance to the non-local attention module with nearly the same testing time and a minimal increase in computational cost.

\begin{figure}[htb!]
\centering
\includegraphics[width=1.00\textwidth]{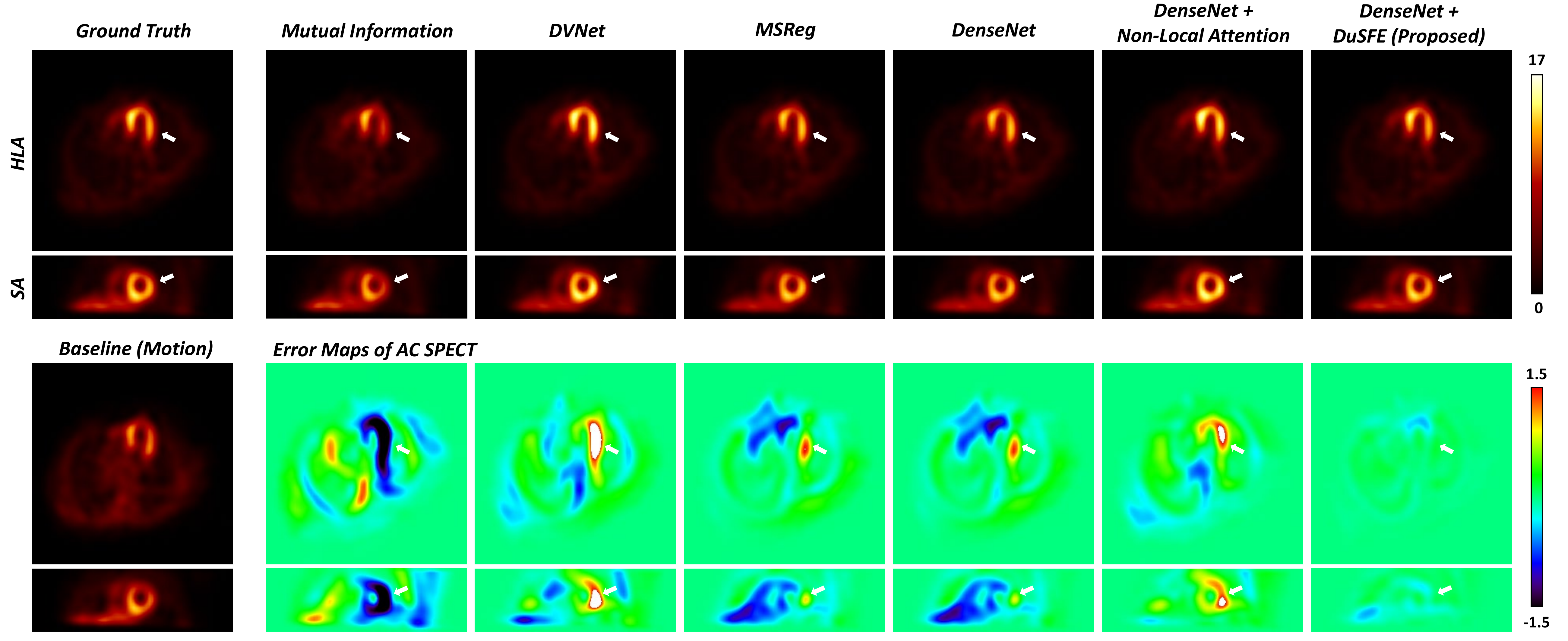}
\caption{Sample AC SPECT images reconstructed using the registered $\mu$-maps. SPECT voxel values are the myocardial perfusion intensities after volume mean normalization.}
\label{fig:ac}
\end{figure}

Fig.~\ref{fig:amap} shows sample registered $\mu$-maps using the predicted registration parameters. Mutual Information showed the highest registration errors. DVNet produced the worst registrations among all deep learning methods. Our proposed DenseNet+DuSFE led to the best results with the lowest registration errors compared to other methods. Fig.~\ref{fig:ac} shows sample AC SPECT images reconstructed using the registered $\mu$-maps. Mutual Information showed the highest activity errors, and DVNet reconstructed the worst results among all deep learning methods. In contrast, our proposed DenseNet+DuSFE generated the most accurate AC SPECT image compared to other methods.

\begin{table} [htb!]
\caption{Quantitative evaluations of registered $\mu$-maps and reconstructed AC SPECT images in terms of NMSE and NMAE. The best results are marked in \textbf{bold}.}
\label{tab:reg_img} 
\scriptsize
\centering
\begin{tabular}{ l | c | c || c | c }
\hline
\multirow{2}{*}{\textbf{Methods}}      & \multicolumn{2}{|c||}{\textbf{Registered} $\boldsymbol{\mu}$\textbf{-maps}}     & \multicolumn{2}{|c}{\textbf{Recon AC SPECT}}                        \Tstrut\Bstrut\\
\cline{2-5}
                                       & \textbf{NMSE($\boldsymbol{\%}$)}   & \textbf{NMAE($\boldsymbol{\%}$)}   & \textbf{NMSE($\boldsymbol{\%}$)}      & \textbf{NMAE($\boldsymbol{\%}$)}    \Tstrut\Bstrut\\
\hline 
Baseline (Motion)                      & $35.81 \pm 13.63$                  & $52.90 \pm 14.23$                  & $23.34 \pm 9.85$                      & $43.81 \pm 10.85$                   \Tstrut\Bstrut\\
Mutual Information \cite{maes1997multimodality}    & $8.59 \pm 7.69$        & $22.52 \pm 10.54$                  & $3.67 \pm 3.53$                       & $17.96 \pm 8.44$                    \Tstrut\Bstrut\\
\hline 
DVNet \cite{sun2018towards}            & $2.94 \pm 2.74$                    & $12.41 \pm 6.12$                   & $1.21 \pm 1.14$                       & $9.94 \pm 4.82$                     \Tstrut\Bstrut\\
MSReg \cite{guo2020deep}               & $2.46 \pm 2.19$                    & $11.36 \pm 5.35$                   & $0.99 \pm 0.88$                       & $9.05 \pm 4.13$                     \Tstrut\Bstrut\\
DenseNet \cite{huang2017densely}       & $2.37 \pm 2.10$                    & $11.14 \pm 5.31$                   & $0.96 \pm 0.84$                       & $8.89 \pm 3.95$                     \Tstrut\Bstrut\\
DenseNet+Non-Local Attention \cite{song2021cross} & $2.09 \pm 1.79$         & $10.55 \pm 4.96$                   & $0.85 \pm 0.70$                       & $8.43 \pm 3.85$                     \Tstrut\Bstrut\\
DenseNet+DuSFE (Proposed)              & $\textbf{1.79} \pm \textbf{1.65}$  & $\textbf{9.54} \pm \textbf{4.90}$  & $\textbf{0.75} \pm \textbf{0.70}$     & $\textbf{7.58} \pm \textbf{4.07}$   \Tstrut\Bstrut\\
\hline 
\end{tabular}
\end{table}

Table~\ref{tab:reg_img} evaluates the registered $\mu$-maps and reconstructed AC SPECT images. Mutual Information output the highest errors. In comparison, our proposed DenseNet+DuSFE showed consistently the lowest errors in both $\mu$-maps and SPECT images among all the methods. Compared to DenseNet+Non-Local Attention, DenseNet+DuSFE generated significantly lower errors in $\mu$-maps ($p$<0.001 for NMSE with paired t-tests) and AC SPECT ($p$=0.003 for NMSE). It demonstrated that DuSFE modules enabled more accurate registration of $\mu$-maps and SPECT, which then improved the accuracy of AC SPECT images.

\section{Conclusion}
We propose a Dual-Branch Squeeze-Fusion-Excitation (DuSFE) module for the cross-modality registration of CT-derived $\mu$-maps and cardiac SPECT images. This study is the first investigation of deep learning methods for the registration of cardiac SPECT/CT. DuSFE effectively fuses the knowledge from the two modalities and then recalibrates both the channel-wise and spatial features for each modality. DuSFE can be further embedded at different convolutional layers to enable gradual feature fusion at different spatial dimensions. In our experiments, a network embedded with DuSFE modules generated more consistent registration parameters with ground truth compared to other traditional and deep learning methods, which then enabled more accurate attenuation correction for cardiac SPECT.

\bibliographystyle{splncs04}
\bibliography{reference}
\end{document}